\pdfoutput=1
\documentclass[12pt]{article}

\usepackage{jcappub}
\usepackage{amsmath}
\usepackage{amssymb}
\usepackage{braket}
\usepackage{graphicx}
\usepackage{hyperref}
\usepackage{mathrsfs}
\usepackage{subcaption}
\usepackage{empheq}
\usepackage{appendix}
\usepackage{natbib}
\usepackage[dvipsnames]{xcolor}
\usepackage{float}
\usepackage[top=1in, bottom=0.4in, left=0.8in, right=0.8in, includefoot]{geometry}

\title{Rescuing Quartic and Natural Inflation \\in the Palatini Formalism}

\author[a,b]{I. Antoniadis,}
\author[c,d]{A. Karam,}
\author[c]{A. Lykkas,}
\author[c]{T. Pappas,}
\author[c]{ and K. Tamvakis}

\affiliation[a]{LPTHE, Sorbonne Universite, CNRS, 4 Place Jussieu, 75005 Paris, France}
\affiliation[b]{Albert Einstein Center, Institute of Theoretical Physics, University of Bern, Sidlerstrasse 5, CH-3012, Bern, Switzerland}
\affiliation[c]{Physics Department, University of Ioannina, GR--45110 Ioannina, Greece}
\affiliation[d]{NICPB, R\"avala 10, 10143 Tallinn, Estonia}

\emailAdd{antoniad@lpthe.jussieu.fr}
\emailAdd{alkaram@cc.uoi.gr}
\emailAdd{alykkas@cc.uoi.gr}
\emailAdd{thpap@cc.uoi.gr}
\emailAdd{tamvakis@uoi.gr}

\abstract{When considered in the Palatini formalism, the Starobinsky model does not provide us with a mechanism for inflation due to the absence of a propagating scalar degree of freedom. By (non)--minimally coupling scalar fields to the Starobinsky model in the Palatini formalism we can in principle describe the inflationary epoch. In this article, we focus on the minimally coupled quartic and natural inflation models. Both theories are excluded in their simplest realization since they predict values for the inflationary observables that are outside the limits set by the Planck data. However, with the addition of the $R^2$ term and the use of the Palatini formalism, we show that these models can be rendered viable.}

\begin{document}

\maketitle

\section{Introduction}

Cosmological inflation~\cite{Starobinsky1980, Sato1981, Guth1981, Linde1982b, Albrecht1982a, Linde1983a, Lyth1999}, namely a phase of de Sitter expansion after the big bang, provides an attractive paradigm addressing the open issues of the early universe. In this framework quantum fluctuations, magnified to a cosmic size, ultimately generate the large scale structure of the universe and the presently observed anisotropy in the CMB. Although a detailed particle physics mechanism underlying inflation has not been established yet, models of inflation formulated in terms of a scalar degree of freedom (inflaton) exist with a number of predictions confirmed by observations. This scalar degree of freedom is introduced either as a fundamental scalar field or can be provided by gravity itself in models featuring generalizations of the Einstein-Hilbert action. The latter possibility is realized in the Starobinsky model where an extra quadratic term of the Ricci curvature scalar $R^2$ is present. Among the models realizing the former possibility of a fundamental scalar, the case of Higgs inflation~\cite{DeSimone2009, Barbon2009, Bezrukov2008a, Barvinsky2008,  Barvinsky2009a, Bezrukov2009a, Lerner2010a, Bezrukov2011, Kamada2012, Bezrukov2013, Hamada2014, Bezrukov2014, Allison2014, Salvio2015, Hamada2015, Calmet2016, Rubio2018, Enckell2018}, where the inflaton is identified with the Standard Model Higgs boson, has attracted a lot of attention since it provides a direct connection of cosmological inflation to particle physics. Higgs inflation requires the presence of an appreciable coupling of the Higgs field $h$ to the Ricci curvature scalar $\xi h^2R$. 

Both of the above possibilities are intimately connected with gravitation and its formulation. It is well known that for the Einstein-Hilbert action of pure gravitation the alternative variational principle of Palatini~\cite{Weyl1950, Deser1976, Hehl1995, Deser2006, Bauer2008}, in which not only the metric $g_{\mu\nu}$ but also the connection $\Gamma_{\mu\nu}^{\rho}$ are treated as independent variables, leads to the standard equations of motion of General Relativity (GR) and the Palatini formulation is equivalent to the standard metric formulation. Nevertheless, this is not the case when extra fields are coupled to gravity non-minimally as in the case of Higgs inflation. New interactions in the scalar sector modifying significantly the Einstein-frame potential are present in the Palatini formulation. The inequivalence of the two formulations is also quite striking in the case of the Starobinsky model, where, in the Palatini formulation there is no propagating extra scalar degree of freedom, in contrast to the metric formulation, where the quadratic curvature term introduces an extra scalar suitable to play the role of the inflaton. 

In the present article we consider scalar matter coupled to gravity in the framework of the Palatini formulation (see~\cite{Sotiriou2010} for a review and~\cite{Bauer2008, Borunda2008, Olmo2011, Bauer2011, Tamanini2011, Enqvist2012, Borowiec2012, Racioppi2017, Rasanen2017, Fu2017, Stachowski2017, Szydlowski2017, Tenkanen2017, Markkanen2018, Carrilho2018, Enckell2018, Enckell2018a, Kozak2018, Jaerv2018, Wang2018, Wu2018, Szydlowski2018, Antoniadis2018, Rasanen2018a, Rasanen2018, Almeida2018} for various applications). We focus on the case that a quadratic Starobinsky term is present while the scalar fields are minimally coupled to gravity\footnote{See~\cite{Cardenas2003, Artymowski2015a, Bruck2015, Asaka2016, Artymowski2016, Kaneda2016, Calmet2016, Bruck2016a, Wang2017, Ema2017, Mori2017, Pi2018, He2018, Gorbunov2018, Ghilencea2018, Wang2018b, Bombacigno2018a, Antoniadis2018, Gundhi2018, Karam2018b} for some alterations of the Starobinsky model with the addition of an extra scalar field.}. We analyze the case of a model with quartic scalar interactions, a case of interest to Higgs inflation. We find acceptable slow-roll inflationary behaviour, despite the fact that scalars are minimally coupled to gravity, contrary to the standard metric case where an appreciable non-minimal coupling is required. The central ingredient of the inflation mechanism operating in the Palatini framework for a model described by a potential $V(\phi)$ is the creation of a plateau for the Einstein-frame potential $\bar{V}(\phi)=V(\phi)/\left(1+4\alpha V(\phi)/M_P^4\right)$ at large field values~\cite{Antoniadis2018, Enckell2018a}. We extend our analysis to the case of the natural inflation model~\cite{Freese1990, Adams1993} characterized by a bounded potential and find that the resulting flattening of the inflationary plateau in the Palatini formulation leads to acceptable slow-roll inflationary behaviour.

The paper is organized as follows: In section~\ref{sec:2} we start with the action of a scalar field minimally coupled to gravity in the presence of a quadratic curvature scalar term with a general self-interacting potential and derive the Einstein frame Lagrangian in the framework of the Palatini formalism. In section~\ref{sec:3.1} we consider the case of a quartic potential modelled in the fashion of the Higgs boson and proceed to analyze the inflationary behaviour calculating the slow-roll parameters and the corresponding inflationary observables. Also, in section~\ref{sec:3.2} we briefly discuss other monomial potentials. Then, in section~\ref{sec:4} we analyze the natural inflation model. Finally, in section~\ref{sec:5} we present our conclusions.

\section{Minimally Coupled Scalars in the Palatini Formalism}
\label{sec:2}

Consider the action of scalar fields minimally coupled to gravity. Although we restrict our consideration to one scalar, the generalization to many is straightforward. We have
\begin{equation} 
{\cal{S}}_0\,=\,\int\,d^4x\,\sqrt{-g}\,\left\{\,\frac{1}{2}M_P^2\,R\,-\frac{1}{2}\left(\nabla\phi\right)^2\,-V(\phi)\,\right\}\ . {\label{ACT-0}}
\end{equation}
We expect that quantum corrections are bound to generate in~\eqref{ACT-0} terms of the form $\xi\phi^2 R$ and $\alpha R^2$. Either of these terms has been shown to lead to a number of interesting results with respect to inflation. It is conceivable though, since  the phenomenological values of the parameters $\xi$ and $\alpha$ are not a priori known, that one or both of these terms are small or negligible. In what follows we shall focus on the case of a negligible non-minimal coupling, i.e. take $\xi=0$, and consider the action
\begin{equation} 
{\cal{S}}\,=\,\int\,d^4x\,\sqrt{-g}\,\left\{\,\frac{1}{2}M_P^2\,R\,+\,\frac{\alpha}{4}R^2\,-\frac{1}{2}\left(\nabla\phi\right)^2\,-V(\phi)\,\right\}\ , {\label{ACT-1}}
\end{equation}
which can be readily written in terms of an auxiliary scalar field $\chi$ as 
\begin{equation}
{\cal{S}}\,=\,\int\,d^4x\,\sqrt{-g}\,\left\{\,\frac{1}{2}M_P^2\left(1+\alpha\chi^2\right)R\,-\frac{\alpha}{4}\chi^4\,-\frac{1}{2}\left(\nabla\phi\right)^2\,-V(\phi)\,\right\}\ .{\label{ACT-2}}
\end{equation}

We shall consider the above action in the framework of the Palatini or first order formalism in which, next to the metric $g_{\mu\nu}$, the connection $\Gamma_{\mu\nu}^{\rho}$ also is an independent variable. Therefore, $R_{\mu\nu}$ is independent of the metric. Performing a Weyl rescaling of the metric according to
 \begin{equation}
\bar{g}_{\mu\nu}\,=\,\frac{1}{M_P^2}\left(M_P^2+\alpha \chi^2\right)\,g_{\mu\nu}\ ,
\end{equation}
we transform the action\eqref{ACT-2} into the Einstein frame
\begin{equation}
{\cal{S}}\,=\,\int\,d^4x\,\sqrt{-\bar{g}}\,\left\{\,\frac{1}{2}M_P^2\,\bar{R}\,-\frac{1}{2}\frac{M_P^2\left(\overline{\nabla}\phi\right)^2}{(M_P^2+\alpha \chi^2)}\,-\overline{V}(\phi,\chi)\,\right\}\ ,{\label{ACT-3}}
\end{equation}
where $\bar{R}=\bar{g}^{\mu\nu}R_{\mu\nu}(\Gamma)$ and
\begin{equation}
\overline{V}(\phi,\chi)\,=\,\frac{M_P^4\,\left(\,V(\phi)\,+\,\frac{\alpha}{4}\chi^4\right)}{\left(M_P^2+\alpha\chi^2\right)^2}\ .
\end{equation}
Variation of the Einstein-Hilbert action~\eqref{ACT-3} with respect to the connection yields the standard Levi-Civita relation in terms of the metric $\bar{g}$. The Einstein field equations are the same as in the standard metric formulation. Varying with respect to the auxiliary field $\chi$ we obtain
\begin{equation}
\frac{\delta{\cal{S}}}{\delta\chi}\,=\,0\,\Longrightarrow\,\chi^2\,=\,\frac{4V(\phi)\,+\,(\nabla\phi)^2}{M_P^2-\alpha\frac{(\nabla\phi)^2}{M_P^2}}\ .{\label{CHI}}
\end{equation}
Note that we have dropped the bars for the sake of simplicity of notation. Substituting~\eqref{CHI} into the action~\eqref{ACT-3} we obtain
\begin{equation}
{\cal{S}}\,=\,\int\,\sqrt{-g}\left\{\,\frac{1}{2}M_P^2\,R\,-\frac{1}{2}\frac{\left(\nabla\phi\right)^2}{\left(1\,+\,\frac{4\alpha}{M_P^4}V(\phi)\,\right)}\,-\frac{V(\phi)}{\left(1\,+\,\frac{4\alpha}{M_P^4}V(\phi)\,\right)}\,+\,O((\nabla\phi)^4)\,\right\}\ .{\label{ACT-4}}
\end{equation}
Since we aim to study the properties of this action in the framework of slow-roll inflation, higher than quadratic powers of $\nabla\phi$ are not expected to play any role. 

The above can be generalized to any $f(R)$ theory replacing the action~\eqref{ACT-3} with
\begin{equation}
{\cal{S}}\,=\,\int\,d^4x\,\sqrt{-g}\,\left\{\,\frac{1}{2}f'(\chi^2)\,R\,-\frac{1}{2}\left(\nabla\phi\right)^2\,-V(\phi)\,-U(\chi^2)\,\right\}\ ,
\end{equation}
where the derivative is with respect to $\chi^2$ and
\begin{equation}
U(\chi^2)\,=\,\frac{1}{2}\left(\,\chi^2\,f'(\chi^2)\,-f(\chi^2)\,\right)\ .{\label{ACT-5}}
\end{equation}
Then, the Weyl rescaling factor is replaced by $\Omega^2=f'(\chi^2)/M_P^2$ and~\eqref{CHI} becomes
\begin{equation}
2f(\chi^2)\,-\chi^2 f'(\chi^2)\,-f'(\chi^2)\frac{(\nabla\phi)^2}{M_P^2}\,=\,4V(\phi)\ .
\end{equation}

\section{Application to Higgs-like Scalars}
\label{sec:3}

\subsection{Quartic potentials}
\label{sec:3.1}

Higgs inflation, i.e. the possibility of the Standard Model Higgs playing the role of the inflaton, has attracted a lot of attention. A necessary requirement of such a scenario within the standard metric formulation is a rather large non-minimal coupling of the Higgs to the Ricci scalar $\xi|H|^2R$. In the unitary gauge the Higgs doublet can be replaced by a single scalar $\phi$
\begin{equation}
H\,=\,\frac{1}{\sqrt{2}}\left(\begin{array}{c}
0\\
\phi\end{array}\right)\,\,\Longrightarrow\,\,\,V(\phi)\,=\,\frac{\lambda}{4}\left(\phi^2-v^2\right)^2\ .{\label{HIGGS-0}}
\end{equation}
Of course, if this is to represent a realistic version of the Standard Model, issues like the stability of the Higgs self-coupling have to be addressed, possibly by enlarging the Higgs sector.

In this section we shall consider the case of a scalar $\phi$ with a quartic potential~\eqref{HIGGS-0} minimally coupled to gravity ($\xi=0$) in the framework of the Palatini formalism. The corresponding scalar part of the Lagrangian derived in~\eqref{ACT-4} is
\begin{equation}
{\cal{L}}(\phi)\,=\,-\frac{1}{2}\frac{(\nabla\phi)^2}{\left(1+\frac{\lambda \alpha}{M_P^4}\left(\phi^2-v^2\right)^2\,\right)}\,-\frac{\frac{\lambda}{4}\left(\phi^2-v^2\right)^2}{\left(1+\frac{\lambda \alpha}{M_P^4}\left(\phi^2-v^2\right)^2\,\right)}\,+\,O((\nabla\phi)^4) \ .{\label{LAG-1}}
\end{equation}
Note that the vanishing of the potential at the (Minkowski) vacuum corresponds through~\eqref{CHI} to a vanishing of the vacuum expectation value of the auxiliary field $\chi$ and consequently to an equality of the Jordan frame and Einstein frame Planck masses.

\begin{figure}[H]
    \centering
    \includegraphics[width=0.5\textwidth]{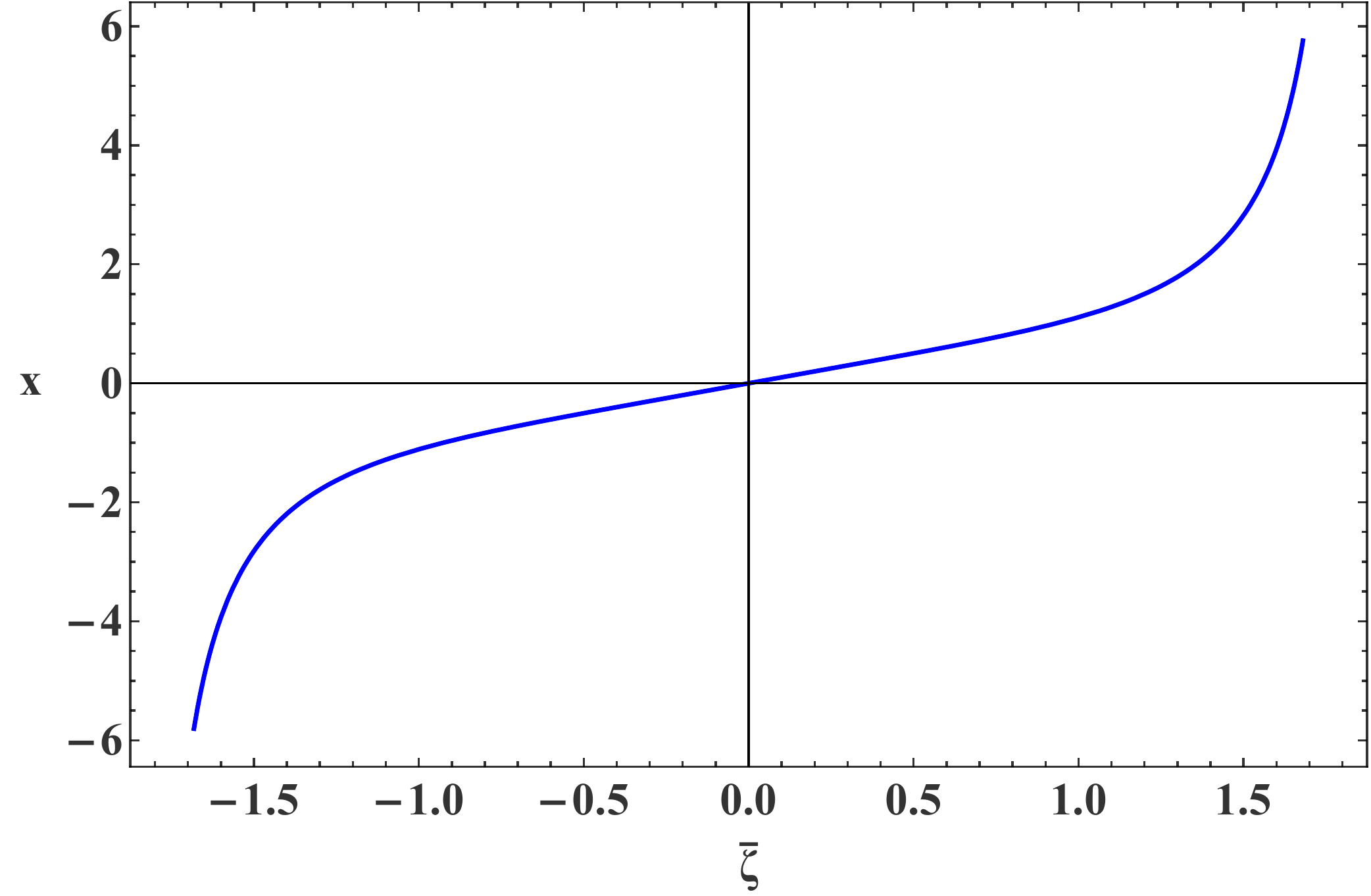}
    \caption{Plot of $x(\bar{\zeta})$.}
    \label{fig:xfz}
\end{figure}

The scalar field $\phi$ can be replaced by a canonically normalized field $\zeta$ defined by
\begin{equation}
\zeta\,=\,\int\,\frac{d\phi}{\sqrt{1+\frac{\lambda \alpha}{M_P^4}\left(\phi^2-v^2\right)^2\,}}\ . {\label{ZETA-0}}
\end{equation}
For very large values of the field $\phi\gg v$ this is approximately
\begin{equation}
    \zeta=M_P\,(\alpha\lambda)^{-1/4}\int\!\frac{\mathrm{d}x}{\sqrt{1+x^4}}\,=\,\frac{M_P}{(\alpha\lambda)^{1/4}}\left(\,\frac{4}{\sqrt{\pi}}\left(\Gamma(5/4)\right)^2\,-\frac{1}{2}{\cal{F}}(y,1/\sqrt{2})\,\right)\,,
\end{equation}
where $x\equiv(\alpha\lambda)^{1/4}\phi/M_P$, $\cos{y}\equiv(x^2-1)/(x^2+1)$ and $\mathcal{F}$ refers to the elliptic integral of the first kind.\footnote{Note that $\frac{4}{\sqrt{\pi}}\left(\Gamma(5/4)\right)^2\,\approx\,1.85407$ and is important in the asymptotic regions of the field $\bar{\zeta}$.} We can invert the above expression and obtain $x(\zeta)$ in terms of one of the Jacobi's elliptic functions ($sn$). In Fig.~\ref{fig:xfz} we present a plot of $x$ in terms of the normalized canonical field $\bar{\zeta}\equiv(\alpha\lambda)^{1/4}\zeta/M_P$.

The $\bar{\zeta}$ function saturates at some value, say $\pm\bar{\zeta}_0$, for $x\rightarrow\pm\infty$. This can also be verified analytically, by means of an asymptotic expansion of $\mathcal{F}$ at large $x$ values.

\begin{figure}[H]
    \centering
    \includegraphics[width=0.55\textwidth]{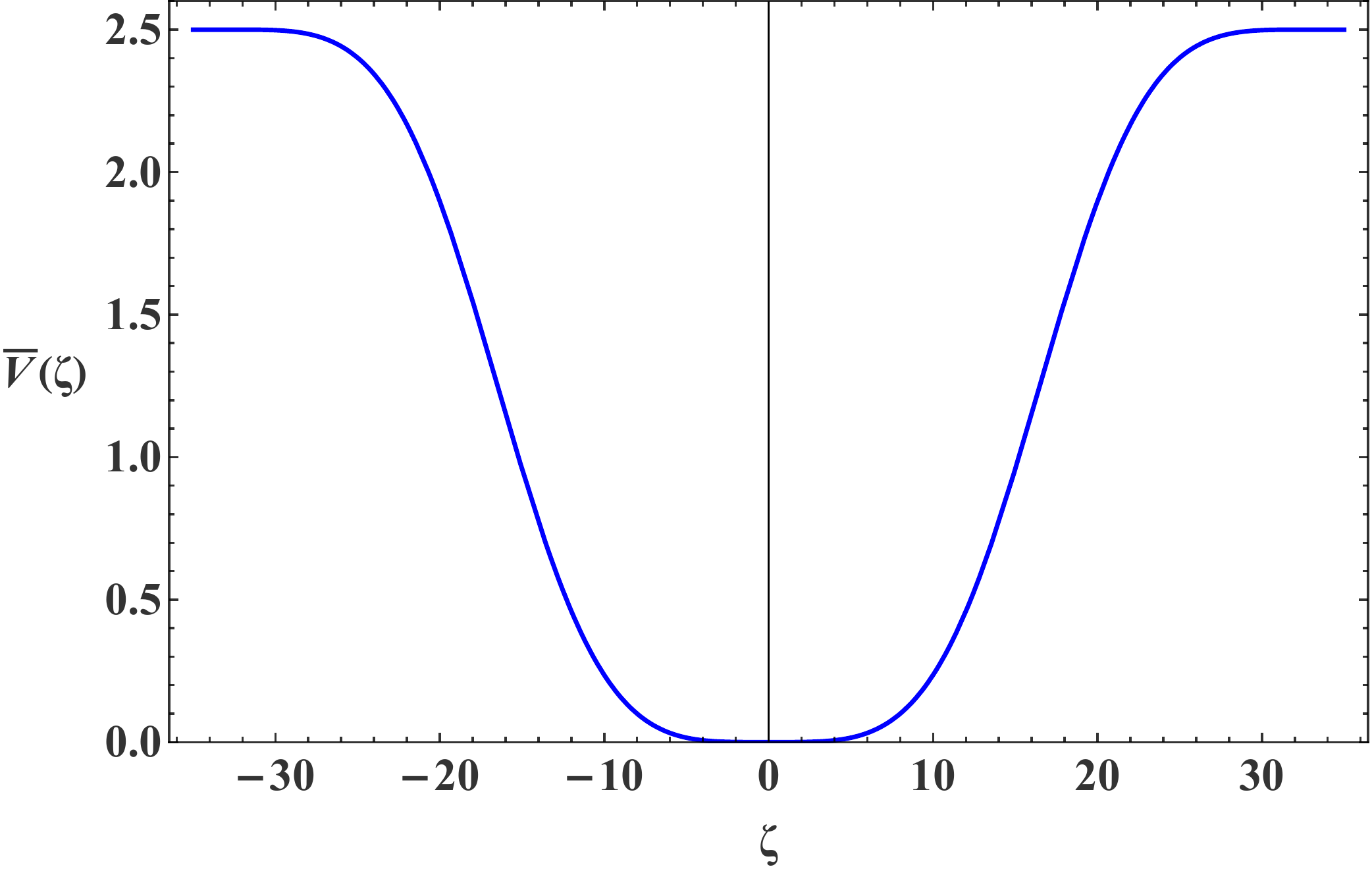}
    \caption{A plot of the potential $\bar{V}(\zeta)$ for $\alpha=0.1$ and $\lambda=10^{-4}$ in natural units.}
    \label{fig:Vz}
\end{figure}

The potential $\bar{V}$ in~\eqref{LAG-1} for the large $\phi\gg v$ region, expressed in terms of $x$, reads
\begin{equation}
    \bar{V}=\frac{M_P^4}{4\alpha}\,\frac{x^4(\zeta)}{\left(1+x^4(\zeta)\right)}\,{\label{POT-1}}
\end{equation}
and it is presented in Fig.~\ref{fig:Vz}. 

Next, we proceed to study slow-roll inflation for this particular model. Substituting the above potential in the slow-roll parameters
\begin{equation}
    \epsilon_V=\frac{M_P^2}{2}\left(\frac{\bar{V}'(\zeta)}{\bar{V}(\zeta)}\right)^2,\qquad \eta_V=M_P^2\left(\frac{\bar{V}''(\zeta)}{\bar{V}(\zeta)}\right)\,,
\end{equation}
we arrive at the expressions
\begin{equation}\label{spar}
    \epsilon_V=\frac{8\sqrt{\alpha\lambda}}{x^2\left(1+x^4\right)}\,,\qquad \eta_V=\frac{12\sqrt{\alpha\lambda}\left(1-x^4\right)}{x^2\left(1+x^4\right)}.
\end{equation}
Then, the number of e-folds can be computed as
\begin{equation}
    N=-\frac{1}{M_P}\int_{\zeta_*}^{\zeta_f}\!\frac{\mathrm{d}\zeta}{\sqrt{2\epsilon_V(\zeta)}}=\frac{1}{M_P}\int_{x_f}^{x_*}\!\frac{\mathrm{d}x}{\sqrt{2\epsilon_V(x)}\sqrt{1+x^4}}=\frac{1}{8\sqrt{\alpha\lambda}}\left(x_*^2-x_f^2\right),{\label{EFOLDS}}
\end{equation}
where $x_*$ and $x_f$ are the values of the $x$ field at the start and end of inflation, respectively. The field value $x_f$ is given by the condition for the end of inflation, i.e. $\epsilon_V\simeq1$, which for $\sqrt{\alpha\lambda}\lesssim10^{-2}$ yields $x_f^2\,\approx\,8\sqrt{\alpha\lambda}$. 

\begin{figure}[H]
    \centering
    \includegraphics[width=0.73\textwidth]{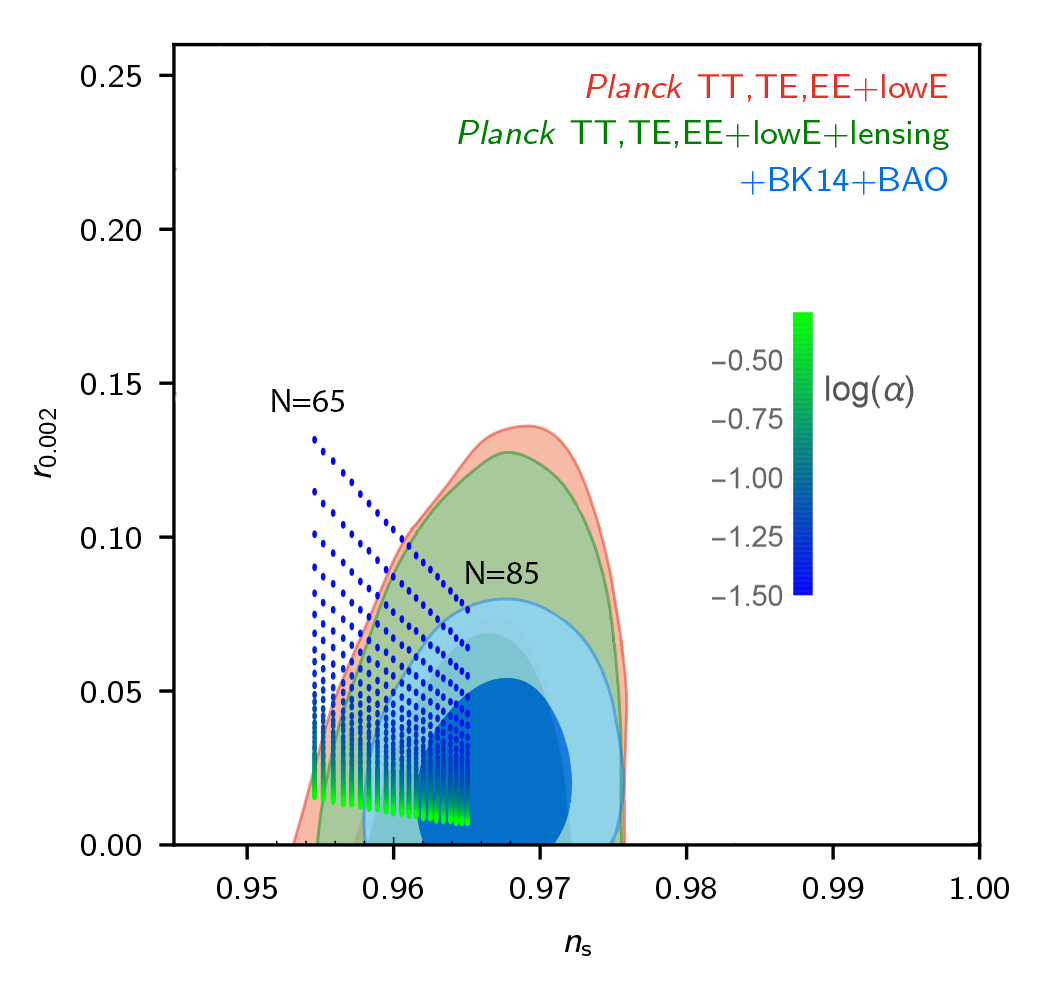}
    \caption{A plot of $r-n_s$. We assumed fixed values of $\lambda\sim10^{-4}$ and $M_P=1$ and varied the $\alpha$ parameter, $\alpha\in(0.01,0.1)$. The number of e-folds represented in this figure are $65-85$.}
    \label{fig:rns}
\end{figure}

Substituting the expressions~\eqref{spar} into the formula for the spectral index $n_s$ we obtain
\begin{equation}
    n_s=1-6\epsilon_V(x_*)+2\eta_V(x_*)\,=\,1-\frac{24\sqrt{\alpha\lambda}}{x_*^2}\,=\,1-\frac{3}{N+x_f^2/8\sqrt{\alpha\lambda}}\,\approx\,\frac{N-2}{N+1}
\end{equation}
and the spectral index turns out to be approximately independent of $\alpha$. This corresponds to a larger than usual number of e-folds, $N\in\left(70,80\right)$, not necessarily out of line with high-scale inflation. This number of e-folds seems to require a period of slower expansion than the standard radiation dominated one. This issue goes beyond the scope of this article and will be investigated in a future work. In Fig.~\ref{fig:rns} we have plotted the tensor-to-scalar ratio $r$ versus the spectral index $n_s$ for $\lambda\sim 10^{-4}$ and $\alpha\in (0.01,\,0.1)$.

Focusing on characteristic values for the parameters, namely $\alpha=0.1$, $\lambda=10^{-4}$, and $N\sim75$, we obtain for the initial and final field values
\begin{align}
    \phi_*\sim25M_P\quad &\text{and}\quad \phi_f\sim3M_P,\nonumber\\
    &\text{or}\\
    \zeta_*\sim19M_P\quad &\text{and}\quad \zeta_f\sim3M_P.\nonumber  
\end{align}
 At the same time the Lyth bound $|\Delta \phi|\gtrsim M_P\sqrt{r/4\pi}$ for this specific field excursion is trivially satisfied.

\begin{figure}[H]
    \centering
    \includegraphics[width=0.74\textwidth]{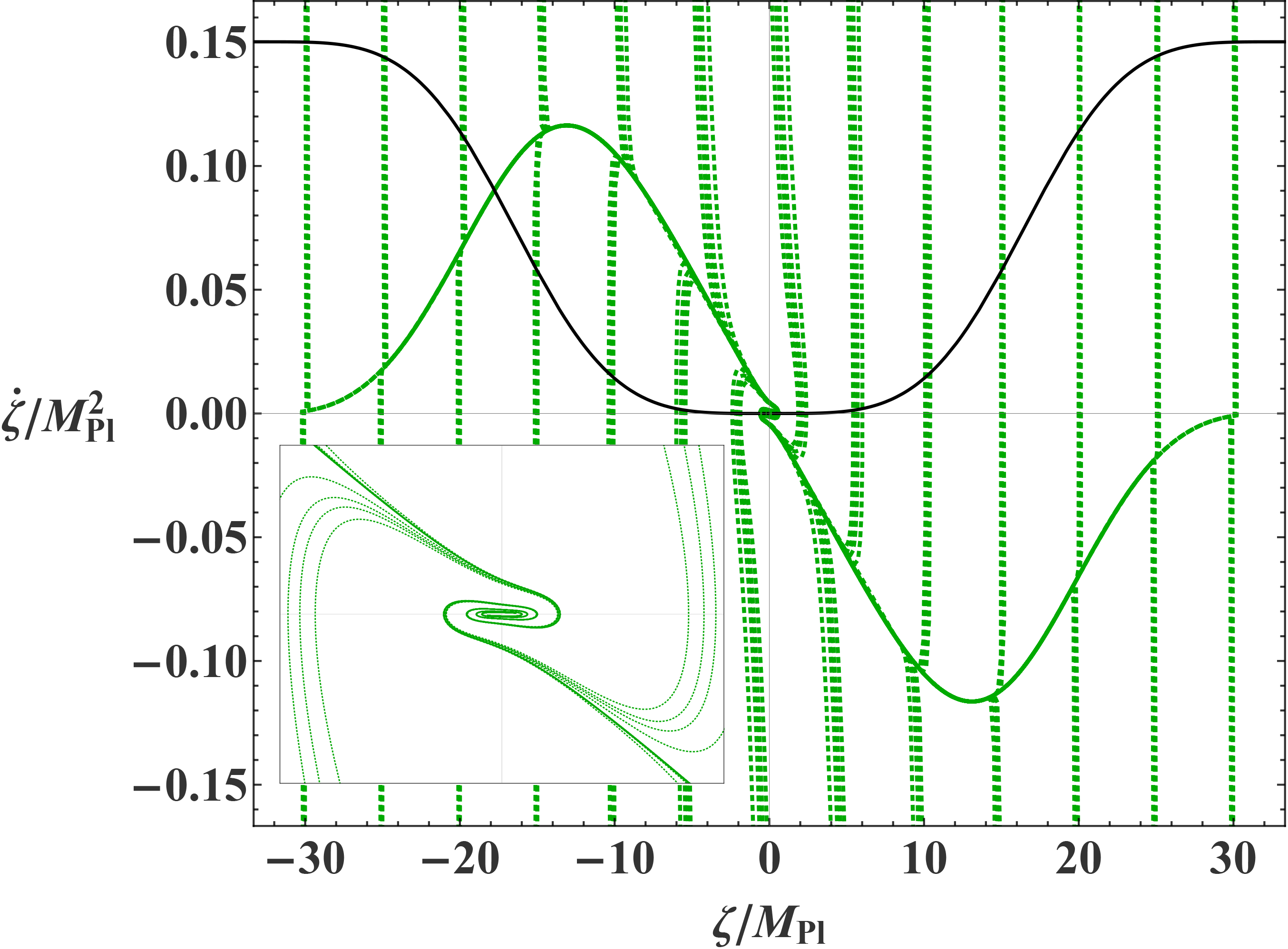}
    \caption{The attractor behaviour of the potential in the $\zeta-\dot{\zeta}$ phase space for $\alpha=0.1$ and $\lambda=10^{-4}$. A magnification of the region close to the minimum of the potential is also included in the bottom left. The black curve corresponds to the normalized potential.}
    \label{fig:Zattractor}
\end{figure}

In Fig.~\ref{fig:Zattractor}, for the fixed value of ${{\alpha=0.1}}$ and $\lambda=10^{-4}$, we solve numerically the Klein-Gordon equation $\ddot{\zeta}+3H\dot{\zeta}+V'(\zeta)=0$ for a plethora of initial conditions for the inflaton $\zeta$ and plot the trajectories in the $\zeta-\dot{\zeta}$ phase space. It is clear that the potential exhibits an attractor behaviour, since, regardless of initial conditions, all trajectories in phase space (green-dotted curves) quickly converge in a single trajectory that ends at the location of the potential minimum. 

\subsection{Other Potentials}
\label{sec:3.2}

It has already been established that a quadratic potential in this Palatini framework leads to an acceptable inflationary behaviour. The quartic potential studied here is also supported by obvious particle physics renormalization and phenomenological arguments. This is not the case for potential functions with a large field behaviour carried out by a power greater than four. Nevertheless, it is interesting to analyze whether this behaviour is shared by more general functions, since for any increasing function $V(\phi)$ the effective potential $\bar{V}(\phi)=V(\phi)/\left(1+\frac{4\alpha}{M_P^4}V(\phi)\right)$ for large values of $\phi$ tends to a plateau $M_P^4/4\alpha$. As a general example we may consider a monomial potential $\lambda\phi^{2n}/4M_P^{2n-4}$. The case $n=1$ (quadratic potential) has been studied elsewhere~\cite{Antoniadis2018} and is proven to exhibit acceptable inflationary behaviour. The integral defining the corresponding canonical field $\zeta$ can be estimated to behave as $C_0-C_1/x^{n-1}$ in terms of $x=(\alpha\lambda)^{\frac{1}{2n}}\phi/M_P$, following an analogous behaviour as in the quartic case. Similarly, the potential $\bar{V}=\frac{M_P^4}{4\alpha}\frac{x^{2n}}{1+x^{2n}}$ as a function of $\zeta$ follows the same behaviour reaching a plateau. The corresponding expressions for the slow-roll parameters are similar, being
\begin{equation}
\epsilon_V\,=\,\frac{2n^2(\alpha\lambda)^{\frac{1}{n}}}{x^2(1+x^{2n})},\qquad\eta_V\,=\,2n(\alpha\lambda)^{\frac{1}{n}}\frac{\left((2n-1)-(n+1)x^{2n}\right)}{x^2\left(1+x^{2n}\right)}\ .
\end{equation}
The value of $x_f^2$ is obtained from $\epsilon_V(x_f)\approx 1$. In the case that $2n^2(\alpha\lambda)^{1/n}$ is small this corresponds to $x_f^2\,\approx\,2n^2(\alpha\lambda)^{1/n}$.
The number of e-folds is
\begin{equation}
N\,=\,\frac{1}{4n(\alpha\lambda)^{\frac{1}{n}}}\left(\,x_*^2\,-x_f^2\,\right)\ .
\end{equation}
Focusing on the $n=3$ case with $\alpha\lambda  \lesssim 10^{-2}$ we have $x_f^2\approx 18(\alpha\lambda)^{1/3}$. Substituting the expressions of $\epsilon_V$, $\eta_V$ for $n=3$ and $x_*^2=12(\alpha \lambda)^{1/3}(N+3/2)$, we obtain for the spectral index
\begin{equation}\label{nsgeneral}
n_s=\left.\frac{2N-(n+2)}{2N+n}\right|_{n=3}=\frac{2N-5}{2N+3}\ ,
\end{equation}
which requires an unacceptably large number of e-folds in order to comply with the observed value of $n_s$. Therefore, the case $n=3$ is excluded for inflation. The same is true for larger values of $n$. Note however that for rational values of $n=q/p$ in the range $1<n<2$ the situation is different. For example $n=3/2$ leads to an acceptable value of $n_s$ for $N=55$. The same is true for $n=2/3$ and $n=4/3$, for $N\approx 50$ and $N\approx 65$ respectively.

General conclusions can also be drawn by taking the large $N$ limit of the above formula \eqref{nsgeneral}
\begin{equation}
n_s=\frac{1-\frac{n+2}{2N}}{1+\frac{n}{2N}}\simeq 1-\frac{n+1}{N}\,,
\end{equation}
which singles out the values of $n$  in the neighbourhood of $n=1$ (quadratic potential), as corresponding to the best fit value $n_s=1-2/N$.

\section{Natural Inflation}
\label{sec:4}

As we saw in the preceding sections the main ingredient of the inflation mechanism operating in the Palatini framework for a model described by a potential $V(\phi)$ is the creation of a plateau for the Einstein-frame potential $\bar{V}(\phi)=V(\phi)/\left(1+4\alpha V(\phi)/M_P^4\right)$ at large field values. This mechanism works perfectly for quadratic and higgs-like quartic potentials leading to acceptable inflationary predictions, although it fails for steeper potential functions. On the other hand, it is possible for a model possessing an inflationary plateau in its standard metric formulation but falling short in its numerical predictions to yield improved results in the Palatini framework. 

As an example, in what follows we consider the so-called natural inflation models, where the role of the inflaton is played by axions, i.e. pseudo-Nambu-Goldstone bosons arising whenever an approximate global symmetry is spontaneously broken~\cite{Freese1990, Adams1993}. We assume a global shift symmetry of the inflaton field is spontaneously broken at some scale $f$, with soft explicit symmetry breaking at a lower scale $M$, which gives the boson its mass. The scalar potential is generally of the form
\begin{equation}
	V(\phi) = M^4 \left( 1 + \cos\left( \frac{\phi}{f} \right) \right) \ .
\end{equation}
The potential has a height $2 M^4$ and a unique minimum at $\phi = \pi f$, assuming the periodicity of $\phi$ is $2\pi f$. For appropriately chosen values of the mass scales, namely, $f \sim M_{P}$ and $M \sim M_{\rm GUT}$ the $\phi$ field can drive inflation. Nevertheless, the latest results from the Planck collaboration have excluded\footnote{However, see Refs\cite{Ferreira2018a, Ferreira2018b} for a possible way to circumvent that in the metric formalism.} natural inflation~\cite{Ade2016, Akrami2018}.

\begin{figure}[H]
    \centering
    \includegraphics[width=0.65\textwidth]{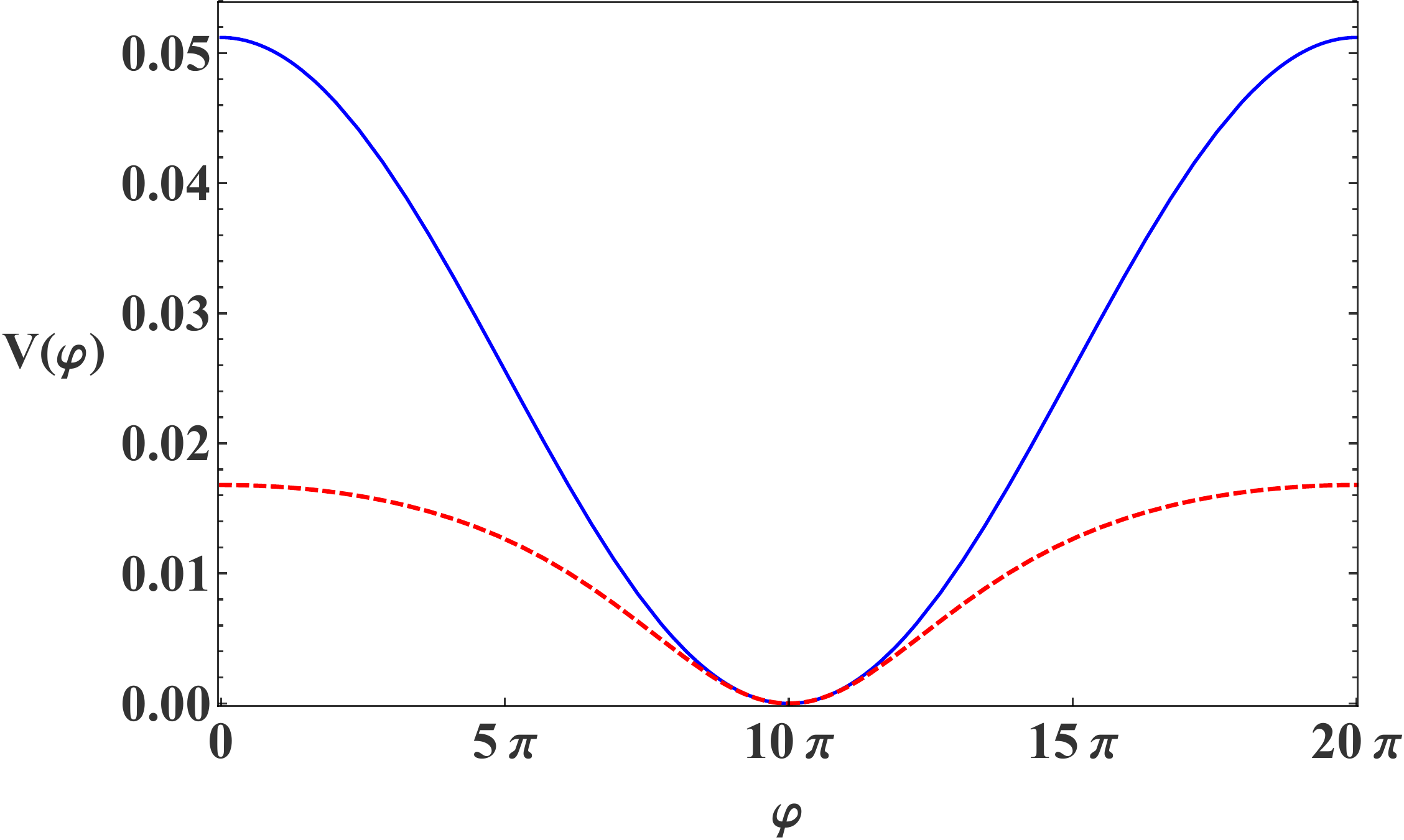}
    \caption{Blue curve: The original natural inflation potential $V(\phi)$. Red dashed curve: The potential $\bar{V}(\zeta(\phi))$ in the Palatini formalism. We have chosen $M = 0.4$, $\alpha = 10$, and $f = 10$ (in natural units).}
    \label{fig:flattening}
\end{figure}

It is interesting to see how the predictions of natural inflation change in the Palatini framework with the $R^2$ term. The integral for the canonical field $\zeta$ yields
\begin{equation}
	\zeta = \frac{2 f}{1 + 2 M^4} {\cal{F}}\left( \phi/2 f, \,2 M^4/(1 + 2 M^4) \right) \ ,
\end{equation}
where ${\cal{F}}$ is again the elliptic integral of the first kind. The above expression can be easily inverted in terms of $\phi$. Then, the effective potential becomes
\begin{equation}
	\bar{V}(\zeta) = \frac{2 M^4\, {\rm cn}^2\left( (1+2M^4)^{1/2}\, \zeta/2f,\,2M^4/(1 + 2M^4) \right)}{1 +  \frac{8\alpha M^4}{M^4_{P}}\,{\rm cn}^2\left((1+2M^4)^{1/2}\, \zeta/2f,\,2M^4/(1 + 2M^4)  \right)  } \ ,
\end{equation}
with ${\rm cn}$ being the Jacobi elliptic function. Choosing $M = 0.5$, $\alpha = 10$, and $f = 10$ (in natural units) we show in Fig.~\ref{fig:flattening} the two potentials $V(\phi)$ and $\bar{V}(\zeta)$. For the chosen values of the parameters, one can see that the potential in the Palatini framework has a flatter plateau. 

Next, we proceed with the calculation of the inflationary observables. We assume slow-roll conditions for the canonically normalized field $\zeta$ and follow a similar line of analysis, as in the previous section. In the following figure we present the $r$-$n_s$ predictions, plotted against the Planck 2018 $1\sigma$ and $2\sigma$ curves~\cite{Akrami2018}.

\begin{figure}[H]
    \centering
    \includegraphics[width=0.75\textwidth]{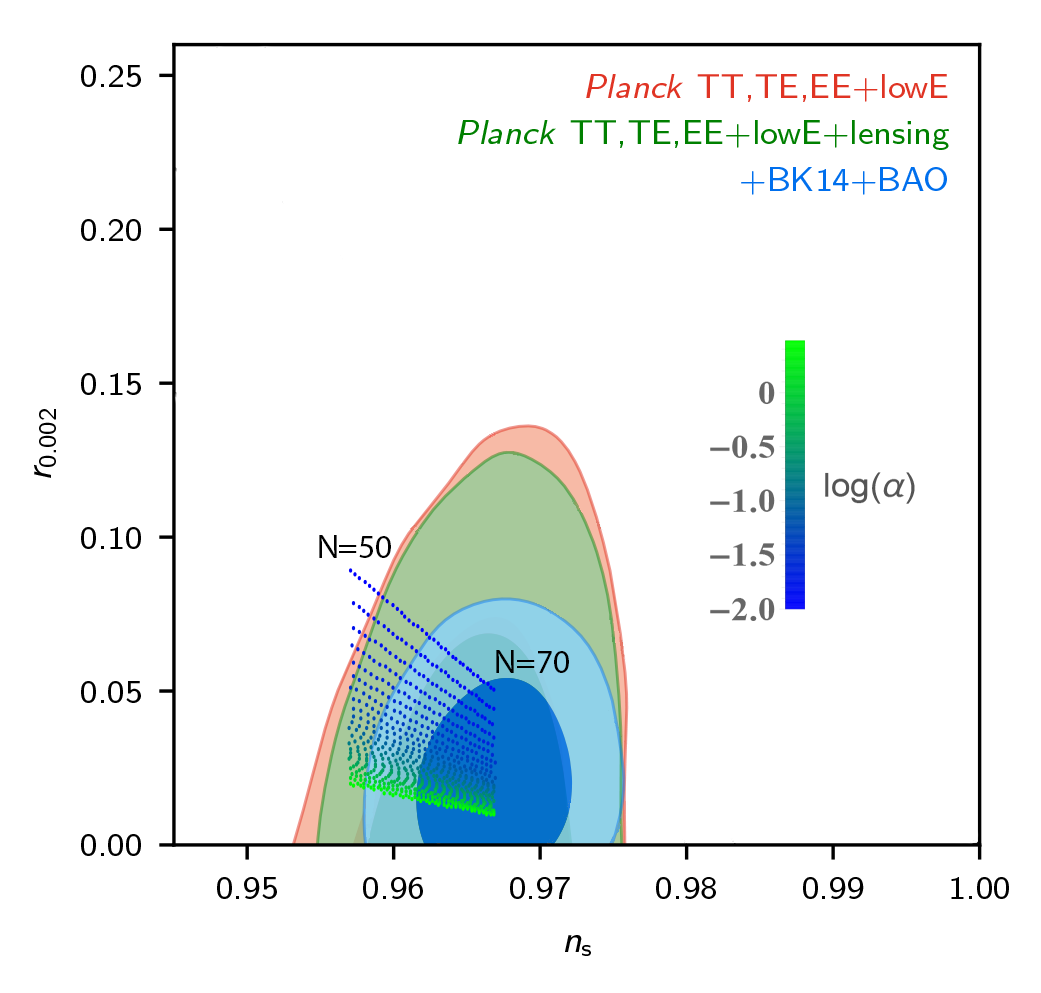}
    \caption{$r-n_s$ plot for $M=0.7, f=7$ and $\alpha \in [0.01,3]$ in natural units.}
    \label{fig:natural_rns}
\end{figure}
For a specific set of the scale parameters $(M,f)$, the variation of $\alpha$ only affects the tensor-to-scalar ratio $r$. By increasing $\alpha$ we obtain smaller values of $r$. There is a lower bound on the scale $f\gtrsim7$ (in natural units), set by the Planck collaboration~\cite{Ade2016}, to which we also abide by here. Increasing the scale $f$ results in pushing the curve to larger values of $r$ and $n_s$.

\begin{figure}[H]
    \centering
    \includegraphics[width=0.7\textwidth]{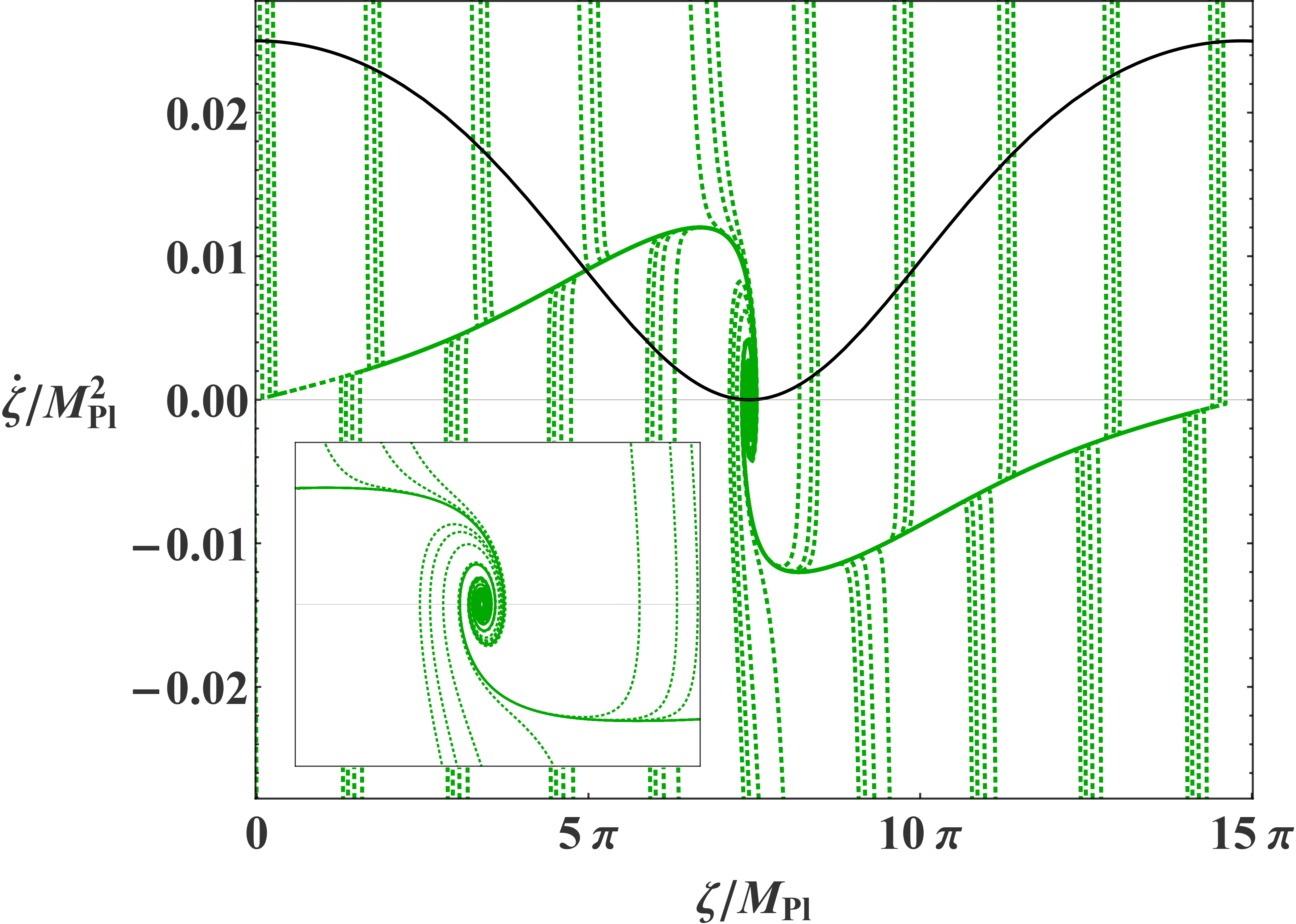}
    \caption{The attractor behaviour of the potential in the $\zeta-\dot{\zeta}$ phase space for $M=0.4 , \alpha=0.01-3$ and $f=7$. A magnification of the region close to the minimum of the potential is also included in the bottom left. The black curve corresponds to the normalized potential.  }
    \label{fig:Attractornatural}
\end{figure}

In Fig.~\ref{fig:Attractornatural}, for the value of ${{\alpha=10}}$ and $f=10$ (the period is $2\pi$), we solve numerically the Klein-Gordon equation $\ddot{\zeta}+3H\dot{\zeta}+\bar{V}'(\zeta)=0$ for a plethora of initial conditions for the inflaton $\zeta$ and plot the trajectories in $\zeta-\dot{\zeta}$ phase space, showing the attractor behaviour of the potential.

\section{Summary and Conclusions}
\label{sec:5}

In this article we considered scalar fields coupled to gravity in the framework of the Palatini formalism. We focused on the case that the theory,  extended with a quadratic curvature term $R/2+\alpha R^2/4$ (rewritten in terms of an auxiliary scalar as $(1+\alpha\chi^2)R/2-\chi^4/4$), also includes fundamental scalar fields coupled minimally to gravity and self-interacting through a scalar potential $V(\phi)$. Transforming the theory to the Einstein frame and integrating out the auxiliary non-propagating scalar degree of freedom we end up with an Einstein frame scalar potential of the form $\overline{V}(\phi)=V(\phi)/\left(1+4\alpha\,V(\phi)\right)$ which, under general conditions, could in principle lead to an inflationary plateau for large field values, this property being a central feature of this framework. 

We analyzed the slow-roll inflationary behaviour in the case of a quartic potential and found acceptable inflationary predictions for the spectral index and tensor to scalar ratio. This case could be of interest for realizing Higgs inflation with a minimal coupling to gravity. We also considered other monomial potentials, although the analogous inflationary behaviour is not shared by steeper potential functions. Next, we analyzed the case of the natural inflation or cosine inflation model, where the role of the inflaton is undertaken by a pseudo-Goldstone boson (axion). Although these models, characterized by a bounded periodic potential, fall short in their inflationary predictions in the standard formulation, when considered in the presence of an $R^2$ term in the framework of the Palatini formalism are shown to yield quite acceptable values for the inflationary observables.

\section*{Acknowledgments}
The research of I.A. is funded in part by the ``Institute Lagrange de Paris", in part by the Swiss National Science Foundation and in part by a CNRS PICS grant. A.K. and T.P. acknowledge support from the Operational Program ``Human Resources Development, Education and Lifelong Learning" which is co-financed by the European Union (European Social Fund) and Greek national funds. A.K. also acknowledges the support of the Estonian Research Council grant MOBJD381 and the ERDF Centre of Excellence project TK133. The research of A.L. is co-financed by Greece and the European Union (European Social Fund - ESF) through the Operational Programme ``Human Resources Development, Education and Lifelong Learning" in the context of the project ``Strengthening Human Resources Research Potential via Doctorate Research'' (MIS-5000432), implemented by the State Scholarships Foundation (IKY).

\bibliography{References}{}

\providecommand{\href}[2]{#2}\begingroup\raggedright\begin{thebibliography}{10}

\bibitem{Starobinsky1980}
A.~A. Starobinsky, {\it {A New Type of Isotropic Cosmological Models Without
  Singularity}},  {\em Phys. Lett.} {\bf 91B} (1980) 99--102.

\bibitem{Sato1981}
K.~Sato, {\it {First Order Phase Transition of a Vacuum and Expansion of the
  Universe}},  {\em Mon. Not. Roy. Astron. Soc.} {\bf 195} (1981) 467--479.

\bibitem{Guth1981}
A.~H. Guth, {\it {The Inflationary Universe: A Possible Solution to the Horizon
  and Flatness Problems}},  {\em Phys. Rev.} {\bf D23} (1981) 347--356.

\bibitem{Linde1982b}
A.~D. Linde, {\it {A New Inflationary Universe Scenario: A Possible Solution of
  the Horizon, Flatness, Homogeneity, Isotropy and Primordial Monopole
  Problems}},  {\em Phys. Lett.} {\bf 108B} (1982) 389--393. [Adv. Ser.
  Astrophys. Cosmol.3,149(1987)].

\bibitem{Albrecht1982a}
A.~Albrecht and P.~J. Steinhardt, {\it {Cosmology for Grand Unified Theories
  with Radiatively Induced Symmetry Breaking}},  {\em Phys. Rev. Lett.} {\bf
  48} (1982) 1220--1223. [Adv. Ser. Astrophys. Cosmol.3,158(1987)].

\bibitem{Linde1983a}
A.~D. Linde, {\it {Chaotic Inflation}},  {\em Phys. Lett.} {\bf 129B} (1983)
  177--181.

\bibitem{Lyth1999}
D.~H. Lyth and A.~Riotto, {\it Particle physics models of inflation and the
  cosmological density perturbation},  {\em Physics Reports} {\bf 314} (1999),
  no.~1 1--146.

\bibitem{DeSimone2009}
A.~De~Simone, M.~P. Hertzberg, and F.~Wilczek, {\it {Running Inflation in the
  Standard Model}},  {\em Phys. Lett.} {\bf B678} (2009) 1--8,
  [\href{http://arxiv.org/abs/0812.4946}{{\tt arXiv:0812.4946}}].

\bibitem{Barbon2009}
J.~L.~F. Barbon and J.~R. Espinosa, {\it {On the Naturalness of Higgs
  Inflation}},  {\em Phys. Rev.} {\bf D79} (2009) 081302,
  [\href{http://arxiv.org/abs/0903.0355}{{\tt arXiv:0903.0355}}].

\bibitem{Bezrukov2008a}
F.~L. Bezrukov and M.~Shaposhnikov, {\it {The Standard Model Higgs boson as the
  inflaton}},  {\em Phys. Lett.} {\bf B659} (2008) 703--706,
  [\href{http://arxiv.org/abs/0710.3755}{{\tt arXiv:0710.3755}}].

\bibitem{Barvinsky2008}
A.~O. Barvinsky, A.~{\relax Yu}. Kamenshchik, and A.~A. Starobinsky, {\it
  {Inflation scenario via the Standard Model Higgs boson and LHC}},  {\em JCAP}
  {\bf 0811} (2008) 021, [\href{http://arxiv.org/abs/0809.2104}{{\tt
  arXiv:0809.2104}}].

\bibitem{Barvinsky2009a}
A.~O. Barvinsky, A.~{\relax Yu}. Kamenshchik, C.~Kiefer, A.~A. Starobinsky, and
  C.~Steinwachs, {\it {Asymptotic freedom in inflationary cosmology with a
  non-minimally coupled Higgs field}},  {\em JCAP} {\bf 0912} (2009) 003,
  [\href{http://arxiv.org/abs/0904.1698}{{\tt arXiv:0904.1698}}].

\bibitem{Bezrukov2009a}
F.~L. Bezrukov, A.~Magnin, and M.~Shaposhnikov, {\it {Standard Model Higgs
  boson mass from inflation}},  {\em Phys. Lett.} {\bf B675} (2009) 88--92,
  [\href{http://arxiv.org/abs/0812.4950}{{\tt arXiv:0812.4950}}].

\bibitem{Lerner2010a}
R.~N. Lerner and J.~McDonald, {\it {A Unitarity-Conserving Higgs Inflation
  Model}},  {\em Phys. Rev.} {\bf D82} (2010) 103525,
  [\href{http://arxiv.org/abs/1005.2978}{{\tt arXiv:1005.2978}}].

\bibitem{Bezrukov2011}
F.~Bezrukov, A.~Magnin, M.~Shaposhnikov, and S.~Sibiryakov, {\it {Higgs
  inflation: consistency and generalisations}},  {\em JHEP} {\bf 01} (2011)
  016, [\href{http://arxiv.org/abs/1008.5157}{{\tt arXiv:1008.5157}}].

\bibitem{Kamada2012}
K.~Kamada, T.~Kobayashi, T.~Takahashi, M.~Yamaguchi, and J.~Yokoyama, {\it
  {Generalized Higgs inflation}},  {\em Phys. Rev.} {\bf D86} (2012) 023504,
  [\href{http://arxiv.org/abs/1203.4059}{{\tt arXiv:1203.4059}}].

\bibitem{Bezrukov2013}
F.~Bezrukov, {\it {The Higgs field as an inflaton}},  {\em Class. Quant. Grav.}
  {\bf 30} (2013) 214001, [\href{http://arxiv.org/abs/1307.0708}{{\tt
  arXiv:1307.0708}}].

\bibitem{Hamada2014}
Y.~Hamada, H.~Kawai, and K.-y. Oda, {\it {Minimal Higgs inflation}},  {\em
  PTEP} {\bf 2014} (2014) 023B02, [\href{http://arxiv.org/abs/1308.6651}{{\tt
  arXiv:1308.6651}}].

\bibitem{Bezrukov2014}
F.~Bezrukov and M.~Shaposhnikov, {\it {Higgs inflation at the critical point}},
   {\em Phys. Lett.} {\bf B734} (2014) 249--254,
  [\href{http://arxiv.org/abs/1403.6078}{{\tt arXiv:1403.6078}}].

\bibitem{Allison2014}
K.~Allison, {\it {Higgs xi-inflation for the 125-126 GeV Higgs: a two-loop
  analysis}},  {\em JHEP} {\bf 02} (2014) 040,
  [\href{http://arxiv.org/abs/1306.6931}{{\tt arXiv:1306.6931}}].

\bibitem{Salvio2015}
A.~Salvio and A.~Mazumdar, {\it {Classical and Quantum Initial Conditions for
  Higgs Inflation}},  {\em Phys. Lett.} {\bf B750} (2015) 194--200,
  [\href{http://arxiv.org/abs/1506.07520}{{\tt arXiv:1506.07520}}].

\bibitem{Hamada2015}
Y.~Hamada, H.~Kawai, K.-y. Oda, and S.~C. Park, {\it {Higgs inflation from
  Standard Model criticality}},  {\em Phys. Rev.} {\bf D91} (2015) 053008,
  [\href{http://arxiv.org/abs/1408.4864}{{\tt arXiv:1408.4864}}].

\bibitem{Calmet2016}
X.~Calmet and I.~Kuntz, {\it {Higgs Starobinsky Inflation}},  {\em Eur. Phys.
  J.} {\bf C76} (2016), no.~5 289, [\href{http://arxiv.org/abs/1605.02236}{{\tt
  arXiv:1605.02236}}].

\bibitem{Rubio2018}
J.~Rubio, {\it {Higgs inflation}},  \href{http://arxiv.org/abs/1807.02376}{{\tt
  arXiv:1807.02376}}.

\bibitem{Enckell2018}
V.-M. Enckell, K.~Enqvist, S.~Rasanen, and E.~Tomberg, {\it {Higgs inflation at
  the hilltop}},  {\em JCAP} {\bf 1806} (2018), no.~06 005,
  [\href{http://arxiv.org/abs/1802.09299}{{\tt arXiv:1802.09299}}].

\bibitem{Weyl1950}
H.~Weyl, {\it {A Remark on the coupling of gravitation and electron}},  {\em
  Phys. Rev.} {\bf 77} (1950) 699--701.

\bibitem{Deser1976}
S.~Deser and C.~J. Isham, {\it {Canonical Vierbein Form of General
  Relativity}},  {\em Phys. Rev.} {\bf D14} (1976) 2505.

\bibitem{Hehl1995}
F.~W. Hehl, J.~D. McCrea, E.~W. Mielke, and Y.~Ne'eman, {\it {Metric affine
  gauge theory of gravity: Field equations, Noether identities, world spinors,
  and breaking of dilation invariance}},  {\em Phys. Rept.} {\bf 258} (1995)
  1--171, [\href{http://arxiv.org/abs/gr-qc/9402012}{{\tt gr-qc/9402012}}].

\bibitem{Deser2006}
S.~Deser, {\it {First-order formalism and odd-derivative actions}},  {\em
  Class. Quant. Grav.} {\bf 23} (2006) 5773,
  [\href{http://arxiv.org/abs/gr-qc/0606006}{{\tt gr-qc/0606006}}].

\bibitem{Bauer2008}
F.~Bauer and D.~A. Demir, {\it {Inflation with Non-Minimal Coupling: Metric
  versus Palatini Formulations}},  {\em Phys. Lett.} {\bf B665} (2008)
  222--226, [\href{http://arxiv.org/abs/0803.2664}{{\tt arXiv:0803.2664}}].

\bibitem{Sotiriou2010}
T.~P. Sotiriou and V.~Faraoni, {\it {f(R) Theories Of Gravity}},  {\em Rev.
  Mod. Phys.} {\bf 82} (2010) 451--497,
  [\href{http://arxiv.org/abs/0805.1726}{{\tt arXiv:0805.1726}}].

\bibitem{Borunda2008}
M.~Borunda, B.~Janssen, and M.~Bastero-Gil, {\it {Palatini versus metric
  formulation in higher curvature gravity}},  {\em JCAP} {\bf 0811} (2008) 008,
  [\href{http://arxiv.org/abs/0804.4440}{{\tt arXiv:0804.4440}}].

\bibitem{Olmo2011}
G.~J. Olmo, {\it {Palatini Approach to Modified Gravity: f(R) Theories and
  Beyond}},  {\em Int. J. Mod. Phys.} {\bf D20} (2011) 413--462,
  [\href{http://arxiv.org/abs/1101.3864}{{\tt arXiv:1101.3864}}].

\bibitem{Bauer2011}
F.~Bauer and D.~A. Demir, {\it {Higgs-Palatini Inflation and Unitarity}},  {\em
  Phys. Lett.} {\bf B698} (2011) 425--429,
  [\href{http://arxiv.org/abs/1012.2900}{{\tt arXiv:1012.2900}}].

\bibitem{Tamanini2011}
N.~Tamanini and C.~R. Contaldi, {\it {Inflationary Perturbations in Palatini
  Generalised Gravity}},  {\em Phys. Rev.} {\bf D83} (2011) 044018,
  [\href{http://arxiv.org/abs/1010.0689}{{\tt arXiv:1010.0689}}].

\bibitem{Enqvist2012}
K.~Enqvist, T.~Koivisto, and G.~Rigopoulos, {\it {Non-metric chaotic
  inflation}},  {\em JCAP} {\bf 1205} (2012) 023,
  [\href{http://arxiv.org/abs/1107.3739}{{\tt arXiv:1107.3739}}].

\bibitem{Borowiec2012}
A.~Borowiec, M.~Kamionka, A.~Kurek, and M.~Szydlowski, {\it {Cosmic
  acceleration from modified gravity with Palatini formalism}},  {\em JCAP}
  {\bf 1202} (2012) 027, [\href{http://arxiv.org/abs/1109.3420}{{\tt
  arXiv:1109.3420}}].

\bibitem{Racioppi2017}
A.~Racioppi, {\it {Coleman-Weinberg linear inflation: metric vs. Palatini
  formulation}},  {\em JCAP} {\bf 1712} (2017), no.~12 041,
  [\href{http://arxiv.org/abs/1710.04853}{{\tt arXiv:1710.04853}}].

\bibitem{Rasanen2017}
S.~Rasanen and P.~Wahlman, {\it {Higgs inflation with loop corrections in the
  Palatini formulation}},  {\em JCAP} {\bf 1711} (2017), no.~11 047,
  [\href{http://arxiv.org/abs/1709.07853}{{\tt arXiv:1709.07853}}].

\bibitem{Fu2017}
C.~Fu, P.~Wu, and H.~Yu, {\it {Inflationary dynamics and preheating of the
  nonminimally coupled inflaton field in the metric and Palatini formalisms}},
  {\em Phys. Rev.} {\bf D96} (2017), no.~10 103542,
  [\href{http://arxiv.org/abs/1801.04089}{{\tt arXiv:1801.04089}}].

\bibitem{Stachowski2017}
A.~Stachowski, M.~Szydłowski, and A.~Borowiec, {\it {Starobinsky cosmological
  model in Palatini formalism}},  {\em Eur. Phys. J.} {\bf C77} (2017), no.~6
  406, [\href{http://arxiv.org/abs/1608.03196}{{\tt arXiv:1608.03196}}].

\bibitem{Szydlowski2017}
M.~Szydłowski, A.~Stachowski, and A.~Borowiec, {\it {Emergence of running dark
  energy from polynomial f(R) theory in Palatini formalism}},  {\em Eur. Phys.
  J.} {\bf C77} (2017), no.~9 603, [\href{http://arxiv.org/abs/1707.01948}{{\tt
  arXiv:1707.01948}}].

\bibitem{Tenkanen2017}
T.~Tenkanen, {\it {Resurrecting Quadratic Inflation with a non-minimal coupling
  to gravity}},  {\em JCAP} {\bf 1712} (2017), no.~12 001,
  [\href{http://arxiv.org/abs/1710.02758}{{\tt arXiv:1710.02758}}].

\bibitem{Markkanen2018}
T.~Markkanen, T.~Tenkanen, V.~Vaskonen, and H.~Veermäe, {\it {Quantum
  corrections to quartic inflation with a non-minimal coupling: metric vs.
  Palatini}},  {\em JCAP} {\bf 1803} (2018), no.~03 029,
  [\href{http://arxiv.org/abs/1712.04874}{{\tt arXiv:1712.04874}}].

\bibitem{Carrilho2018}
P.~Carrilho, D.~Mulryne, J.~Ronayne, and T.~Tenkanen, {\it {Attractor Behaviour
  in Multifield Inflation}},  {\em JCAP} {\bf 1806} (2018), no.~06 032,
  [\href{http://arxiv.org/abs/1804.10489}{{\tt arXiv:1804.10489}}].

\bibitem{Enckell2018a}
V.-M. Enckell, K.~Enqvist, S.~Rasanen, and L.-P. Wahlman, {\it {Inflation with
  $R^2$ term in the Palatini formalism}},
  \href{http://arxiv.org/abs/1810.05536}{{\tt arXiv:1810.05536}}.

\bibitem{Kozak2018}
A.~Kozak and A.~Borowiec, {\it {Palatini frames in scalar-tensor theories of
  gravity}},  \href{http://arxiv.org/abs/1808.05598}{{\tt arXiv:1808.05598}}.

\bibitem{Jaerv2018}
L.~J\"{a}rv, A.~Racioppi, and T.~Tenkanen, {\it {Palatini side of inflationary
  attractors}},  {\em Phys. Rev.} {\bf D97} (2018), no.~8 083513,
  [\href{http://arxiv.org/abs/1712.08471}{{\tt arXiv:1712.08471}}].

\bibitem{Wang2018}
Z.~Wang, P.~Wu, and H.~Yu, {\it {Stability analysis for non-minimally coupled
  dark energy models in the Palatini formalism}},  {\em Astrophys. Space Sci.}
  {\bf 363} (2018), no.~6 120.

\bibitem{Wu2018}
J.~Wu, G.~Li, T.~Harko, and S.-D. Liang, {\it {Palatini formulation of $f(R,T)$
  gravity theory, and its cosmological implications}},  {\em Eur. Phys. J.}
  {\bf C78} (2018), no.~5 430, [\href{http://arxiv.org/abs/1805.07419}{{\tt
  arXiv:1805.07419}}].

\bibitem{Szydlowski2018}
M.~Szydłowski and A.~Stachowski, {\it {Polynomial $f(R)$ Palatini cosmology --
  dynamical system approach}},  {\em Phys. Rev.} {\bf D97} (2018), no.~10
  103524, [\href{http://arxiv.org/abs/1712.00822}{{\tt arXiv:1712.00822}}].

\bibitem{Antoniadis2018}
I.~Antoniadis, A.~Karam, A.~Lykkas, and K.~Tamvakis, {\it {Palatini inflation
  in models with an $R^2$ term}},  {\em JCAP} {\bf 1811} (2018), no.~11 028,
  [\href{http://arxiv.org/abs/1810.10418}{{\tt arXiv:1810.10418}}].

\bibitem{Rasanen2018a}
S.~Rasanen and E.~Tomberg, {\it {Planck scale black hole dark matter from Higgs
  inflation}},  \href{http://arxiv.org/abs/1810.12608}{{\tt arXiv:1810.12608}}.

\bibitem{Rasanen2018}
S.~Rasanen, {\it {Higgs inflation in the Palatini formulation with kinetic
  terms for the metric}},  \href{http://arxiv.org/abs/1811.09514}{{\tt
  arXiv:1811.09514}}.

\bibitem{Almeida2018}
J.~P.~B. Almeida, N.~Bernal, J.~Rubio, and T.~Tenkanen, {\it {Hidden Inflaton
  Dark Matter}},  \href{http://arxiv.org/abs/1811.09640}{{\tt
  arXiv:1811.09640}}.

\bibitem{Cardenas2003}
V.~H. Cardenas, S.~del Campo, and R.~Herrera, {\it {R**2 corrections to chaotic
  inflation}},  {\em Mod. Phys. Lett.} {\bf A18} (2003) 2039--2050,
  [\href{http://arxiv.org/abs/gr-qc/0308040}{{\tt gr-qc/0308040}}].

\bibitem{Artymowski2015a}
M.~Artymowski, Z.~Lalak, and M.~Lewicki, {\it {Inflationary scenarios in
  Starobinsky model with higher order corrections}},  {\em JCAP} {\bf 1506}
  (2015) 032, [\href{http://arxiv.org/abs/1502.01371}{{\tt arXiv:1502.01371}}].

\bibitem{Bruck2015}
C.~van~de Bruck and L.~E. Paduraru, {\it {Simplest extension of Starobinsky
  inflation}},  {\em Phys. Rev.} {\bf D92} (2015) 083513,
  [\href{http://arxiv.org/abs/1505.01727}{{\tt arXiv:1505.01727}}].

\bibitem{Asaka2016}
T.~Asaka, S.~Iso, H.~Kawai, K.~Kohri, T.~Noumi, and T.~Terada, {\it
  {Reinterpretation of the Starobinsky model}},  {\em PTEP} {\bf 2016} (2016),
  no.~12 123E01, [\href{http://arxiv.org/abs/1507.04344}{{\tt
  arXiv:1507.04344}}].

\bibitem{Artymowski2016}
M.~Artymowski, Z.~Lalak, and M.~Lewicki, {\it {Saddle point inflation from
  higher order corrections to Higgs/Starobinsky inflation}},  {\em Phys. Rev.}
  {\bf D93} (2016), no.~4 043514, [\href{http://arxiv.org/abs/1509.00031}{{\tt
  arXiv:1509.00031}}].

\bibitem{Kaneda2016}
S.~Kaneda and S.~V. Ketov, {\it {Starobinsky-like two-field inflation}},  {\em
  Eur. Phys. J.} {\bf C76} (2016), no.~1 26,
  [\href{http://arxiv.org/abs/1510.03524}{{\tt arXiv:1510.03524}}].

\bibitem{Bruck2016a}
C.~van~de Bruck, P.~Dunsby, and L.~E. Paduraru, {\it {Reheating and preheating
  in the simplest extension of Starobinsky inflation}},
  \href{http://arxiv.org/abs/1606.04346}{{\tt arXiv:1606.04346}}.

\bibitem{Wang2017}
Y.-C. Wang and T.~Wang, {\it {Primordial perturbations generated by Higgs field
  and $R^2$ operator}},  {\em Phys. Rev.} {\bf D96} (2017), no.~12 123506,
  [\href{http://arxiv.org/abs/1701.06636}{{\tt arXiv:1701.06636}}].

\bibitem{Ema2017}
Y.~Ema, {\it {Higgs Scalaron Mixed Inflation}},  {\em Phys. Lett.} {\bf B770}
  (2017) 403--411, [\href{http://arxiv.org/abs/1701.07665}{{\tt
  arXiv:1701.07665}}].

\bibitem{Mori2017}
T.~Mori, K.~Kohri, and J.~White, {\it {Multi-field effects in a simple
  extension of $R^2$ inflation}},  {\em JCAP} {\bf 1710} (2017), no.~10 044,
  [\href{http://arxiv.org/abs/1705.05638}{{\tt arXiv:1705.05638}}].

\bibitem{Pi2018}
S.~Pi, Y.-l. Zhang, Q.-G. Huang, and M.~Sasaki, {\it {Scalaron from
  $R^2$-gravity as a heavy field}},  {\em JCAP} {\bf 1805} (2018), no.~05 042,
  [\href{http://arxiv.org/abs/1712.09896}{{\tt arXiv:1712.09896}}].

\bibitem{He2018}
M.~He, A.~A. Starobinsky, and J.~Yokoyama, {\it {Inflation in the mixed
  Higgs-$R^2$ model}},  {\em JCAP} {\bf 1805} (2018), no.~05 064,
  [\href{http://arxiv.org/abs/1804.00409}{{\tt arXiv:1804.00409}}].

\bibitem{Gorbunov2018}
D.~Gorbunov and A.~Tokareva, {\it {Scalaron the healer: removing the
  strong-coupling in the Higgs- and Higgs-dilaton inflations}},
  \href{http://arxiv.org/abs/1807.02392}{{\tt arXiv:1807.02392}}.

\bibitem{Ghilencea2018}
D.~M. Ghilencea, {\it {Two-loop corrections to Starobinsky-Higgs inflation}},
  \href{http://arxiv.org/abs/1807.06900}{{\tt arXiv:1807.06900}}.

\bibitem{Wang2018b}
S.-J. Wang, {\it {Quintessential Starobinsky inflation and swampland
  criteria}},  \href{http://arxiv.org/abs/1810.06445}{{\tt arXiv:1810.06445}}.

\bibitem{Bombacigno2018a}
F.~Bombacigno and G.~Montani, {\it {Cosmological implementation of Palatini
  $f(R)$ models with a Nieh-Yan term}},
  \href{http://arxiv.org/abs/1809.07563}{{\tt arXiv:1809.07563}}.

\bibitem{Gundhi2018}
A.~Gundhi and C.~F. Steinwachs, {\it {Scalaron-Higgs inflation}},
  \href{http://arxiv.org/abs/1810.10546}{{\tt arXiv:1810.10546}}.

\bibitem{Karam2018b}
A.~Karam, T.~Pappas, and K.~Tamvakis, {\it {Nonminimal Coleman--Weinberg
  Inflation with an $R^2$ term}},  \href{http://arxiv.org/abs/1810.12884}{{\tt
  arXiv:1810.12884}}.

\bibitem{Freese1990}
K.~Freese, J.~A. Frieman, and A.~V. Olinto, {\it {Natural inflation with pseudo
  - Nambu-Goldstone bosons}},  {\em Phys. Rev. Lett.} {\bf 65} (1990)
  3233--3236.

\bibitem{Adams1993}
F.~C. Adams, J.~R. Bond, K.~Freese, J.~A. Frieman, and A.~V. Olinto, {\it
  {Natural inflation: Particle physics models, power law spectra for large
  scale structure, and constraints from COBE}},  {\em Phys. Rev.} {\bf D47}
  (1993) 426--455, [\href{http://arxiv.org/abs/hep-ph/9207245}{{\tt
  hep-ph/9207245}}].

\bibitem{Ferreira2018a}
R.~Z. Ferreira and A.~Notari, {\it {Thermalized axion inflation: natural and
  monomial inflation with small $r$}},  {\em Phys. Rev.} {\bf D97} (2018),
  no.~6 063528, [\href{http://arxiv.org/abs/1711.07483}{{\tt
  arXiv:1711.07483}}].

\bibitem{Ferreira2018b}
R.~Z. Ferreira, A.~Notari, and G.~Simeon, {\it {Natural Inflation with a
  periodic non-minimal coupling}},  {\em JCAP} {\bf 1811} (2018), no.~11 021,
  [\href{http://arxiv.org/abs/1806.05511}{{\tt arXiv:1806.05511}}].

\bibitem{Ade2016}
{\bf Planck} Collaboration, P.~A.~R. Ade et~al., {\it {Planck 2015 results. XX.
  Constraints on inflation}},  {\em Astron. Astrophys.} {\bf 594} (2016) A20,
  [\href{http://arxiv.org/abs/1502.02114}{{\tt arXiv:1502.02114}}].

\bibitem{Akrami2018}
{\bf Planck} Collaboration, Y.~Akrami et~al., {\it {Planck 2018 results. X.
  Constraints on inflation}},  \href{http://arxiv.org/abs/1807.06211}{{\tt
  arXiv:1807.06211}}.

\end{thebibliography}\endgroup
\bibliographystyle{jhep}

\end{document}